**First-principles investigation of the emergence of multiferroicity and skyrmions in CrI$_2$**

Khimananda Acharya[1], Kai Huang[2], Evgeny Y. Tsymbal[2] and Tula R. Paudel[1]

[1]Department of Physics, South Dakota School of Mines and Technology, Rapid City, SD 57701, USA

[2]Department of Physics and Astronomy & Nebraska Center for Materials and Nanoscience, University of Nebraska, Lincoln, Nebraska 68588-0299, USA

**Abstract**

The recently discovered van der Waals magnetic semiconductor CrI$_2$ shows promise for spintronic applications. Its electronic and magnetic properties are known to be strongly influenced by electronic correlations. In this work, we employ density functional theory calculations where electronic correlations in CrI$_2$ are considered within an on-site Coulomb interaction, U, with U being determined using the perturbative approach based on the linear response method. We show that the accuracy of an on-site Coulomb interaction is essential for predicting the non-centrosymmetric orthorhombic ground state of CrI$_2$, which allows the existence of ferroelectricity and non-zero Dzyaloshinskii-Moriya (DMI) interaction. Our calculation shows that the ground state of bulk CrI$_2$ has electric polarization of 0.15 $\mu$C/cm$^{-2}$ pointing along the z-axis and the small DMI energy that changes sign with the ferroelectric polarization switching. The DMI and polarization are enhanced to 0.28 meV/formula-unit and 0.63 $\mu$C/cm$^{-2}$ when Pt intercalates bilayers of CrI$_2$, due to its large spin-orbit coupling strength and large off-center displacement. Such an enhanced DMI leads to the Néel skyrmions, whose handedness is controlled by ferroelectric polarization. Our work contributes to the creation and manipulation of bits in skyrmions-based memory devices.

**1. Introduction**

Binary phases of chromium and iodine include CrI$_2$ and CrI$_3$. While CrI$_3$ has generated massive interest due to controllable layer-dependent magnetism[1–3] enabling ultrathin spintronics devices such as magnetic tunnel junctions[4] and spin valve transistors[5], CrI$_2$ has remained understudied and less understood. Unlike CrI$_3$, for which a bulk crystal is available commercially, CrI$_2$ is not available, though a single crystal was reported back in 1973[6]. Noticeably, there are far fewer studies relating to CrI$_2$ than CrI$_3$, including just a few on CrI$_2$ growth: co-deposition of Cr and I$_2$ atoms on graphitized 6H-SiC (0001)[7], molecular beam epitaxy growth Si [111] surface[8] along with Knudsen effusion techniques[9].

Bulk van der Waals (vdW) CrI$_2$ crystallizes in a monoclinic structure called *M* phase, or an orthorhombic crystal structure called *O* phase (Fig. 1). In the *O* phase, CrI$_2$ contains two monolayers in the unit cell, where the top layer can be obtained from the bottom layer by 180° rotation along the c-axis and translation of one layer with respect to other. This stacking order breaks the inversion symmetry, thereby enabling ferroelectricity in the material.[10] The *O* phase CrI$_2$ is also magnetic, as ferromagnetism[7] and helimagnetism[11] been reported, which confirms a multiferroic nature of this material.

The presence of Cr$^{2+}$ ion with $d^{4+}$ electronic configuration and small hybridization with I-*p* states result in large local moments and strong electronic correlations, which are difficult to take into account accurately. The resulting difficulties have led to different theoretical studies reporting different ground states [7,10,12–15]. Typically, in the first principles density functional theory (DFT) based calculations, electronic correlations are treated at the level of the on-site Columb interaction, Hubbard $U$. Non-spin polarized calculations without $U$ predict CrI$_2$ to be metallic[7], contradicting the scanning tunneling spectroscopy band gap measurement of ~3.2 eV [7]. Spin-polarized calculations, on the other hand, report an insulating nature. However, the nature of magnetism, collinear vs. non-collinear, and ferromagnetic (FM) vs. antiferromagnetic (AFM) depends on the magnitude of $U$[7]. Earlier reports have shown that the predicted ground state phases switch from frustrated/canted AFM to FM when $U > 5.0$ eV.[7] This sensitivity to electronic correlations in DFT calculations makes accurately determination of $U$ critically important.

Knowledge of the ground state properties of CrI$_2$ is important for predicting the emergence of topological magnetic quasiparticles in this material. The O phase CrI$_2$ is a promising candidate to realize magnetic skyrmions. The absence of inversion symmetry and the presence of spin orbit coupling (SOC) causes an emergent Dzyaloshinskii–Moriya interaction (DMI),[16] which favors a spin canting between neighboring magnetic atoms. This interaction, along with the Heisenberg interaction and magnetic anisotropy, can lead to the formation of magnetic skyrmions.[17,18].

In this work, we use DFT calculations and the linear response method to determine the on-site Coulomb interaction to determine phase stability, magnetism, and multiferroicity of bulk and mono/few layers of CrI$_2$. We show that multiferroic nature of this material combined with the presence of non-zero DMI, which, when enhanced by intercalation of high spin-orbit coupling element, Pt, leads to skyrmions-like topological magnetic textures with handedness controlled by ferroelectric polarization. Our work contributes to the electrical control of magnetic signals, an important step for spintronics devices.

## 2. Computational Details

CrI$_2$ is studied using the spin-polarized DFT method implemented in the Vienna ab initio simulation package (VASP) [19,20]. The electron-ion potential is approximated using the projected augmented plane wave method (PAW)[21]. The exchange and correlation potential are included using the generalized density approximation (GGA). In these calculations, we use a kinetic energy cutoff of 320 eV for the plane wave expansion of the PAWs and a 6 × 6 × 2 grid of Γ centered $k$-points [22] for the Brillouin zone integration. The exchange and correlation, beyond GGA, are considered by introducing an on-site Coulomb repulsion with Hubbard $U$ = 7.0 eV for Cr 3d orbitals in rotationally invariant formalism.[23] This value of U is obtained using the perturbation-based approach based on the linear response method[24]. The vdW interaction is included in the calculations to describe the interlayer separation using the DFT-D3 method.[25] The unit cell is optimized until the forces on all atoms were less than 0.01 eV using conjugate gradient algorithms. The convergence of calculation with respect to energy cut-off and $k$-points is confirmed.

The phonon band structure calculation is performed using a 4 × 4 × 1 supercell with a finite displacement approach using Phonopy code[26]. Atoms displacement from equilibrium is set to 0.01 Å. The *ab-initio* molecular dynamics simulation is performed using a 3 × 3 × 1 supercell at 300 K in an NVT ensemble using a Nose-Hoover thermostat as implemented in VASP.

Micromagnetic simulation is carried out using Ubermag [27] where OOMMF [28] is used as a numerical calculator.

## 3. Results

### 3.1. Crystal Structure

Fig. 1(c) shows the monoclinic, *M* phase[29] $CrI_2$ with the space group C2/m, and Fig. 1(d) shows the orthorhombic, *O* phase[30] $CrI_2$ with the space group $Cmc2_1$. These two structures have nearly equal Cr-Cr distance, and hence are difficult to distinguish in STM imaging[8]. The evaluated lattice parameters for the minimum energy structures, listed in Tables S1 and S2 (Supplemental Material) compare well with experimental parameters. In the *M* phase, layers are stacked on each other; two penultimate layers are related by mirror reflection maintaining inversion symmetry. In contrast, the *O* phase is produced by displacement along $\hat{b}$ direction for 0.653*b* followed by reflection between the layers resulting in a structure with broken inversion symmetry. Notably, both phases appear identical in a single-layer limit as shown in Fig. 1(a). Within the single layer, the Cr layer is sandwiched between the two I layers so that a Cr atom is bonded to six I atoms, forming a tilted octahedron like that in other transition metal dichalcogenides. The distances between chromium and iodine along different diagonals are slightly different resulting in distorted octahedra.

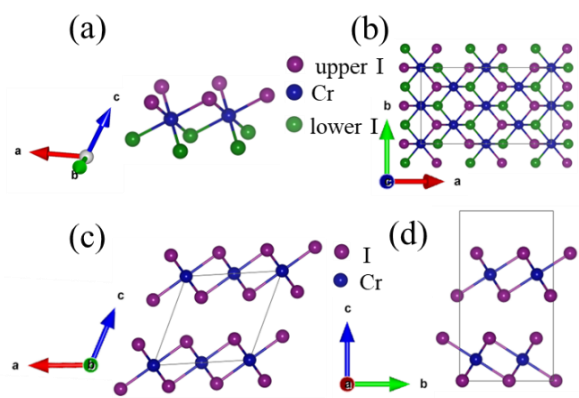

Figure 1: (a) 3D view of a single layer of $CrI_2$ showing Cr atoms in I octahedra, green(pink) represents the lower(upper) I atoms. (b) Top view of monolayer $CrI_2$. (c,d) Side view of *M* phase (c), and *O* phase (d) $CrI_2$.

## 3.2. Electronic Correlations

In our DFT calculations, electronic correlations are treated at the level of the on-site Columb interaction, Hubbard U. We determine the U using the perturbation-based approach based on the linear response method.[24] In this approach, we evaluate the difference in occupation of Cr $3d$- orbitals with (SCF) and without (NSCF) screening when a small spherical perturbations $V_I$ is introduced to Cr 3d orbitals, i.e.

$$U = \left(\frac{\partial N_I^{SCF}}{\partial V_I}\right)^{-1} - \left(\frac{\partial N_I^{NSCF}}{\partial V_I}\right)^{-1}.$$

Fig. 2(a) shows evaluated values $\left(\frac{\partial N_I^{SCF}}{\partial V_I}\right)^{-1}$ and $\left(\frac{\partial N_I^{SCF}}{\partial V_I}\right)^{-1}$ plotted as a function of applied potential $V_I$. We use small perturbation $V_I < 0.1$ eV to ensure the validity of the linear response theory. The $U$ is then determined from the difference in slope resulting in 7.1 eV for the *M* phase and 6.9 eV for the *O* phase. These values are rather large compared to the typical values of 3-5 eV assumed in literature [3,31–33]. Fig. 2(b) shows the convergence of $U$ with the number of Cr atoms. It converges rather slowly compared to 3D crystals like NiO[24] which is used in the original work employing the linear response method. The large value of $U$, however, can be understood in terms of the 2D charge localization and strong screening. The evaluated charge distribution for 3D crystal NiO and CrI$_2$ is markedly different, as shown in Fig. 2(c, d). In contrast to 3D response charge distribution (resulting from small perturbation applied to Ni 3d orbitals) that spans beyond 3$^{rd}$ nearest neighbors in NiO, it is purely 2D, merely spans to the first nearest neighbor with almost entirely focused on the perturbed Cr atom itself in CrI$_2$.

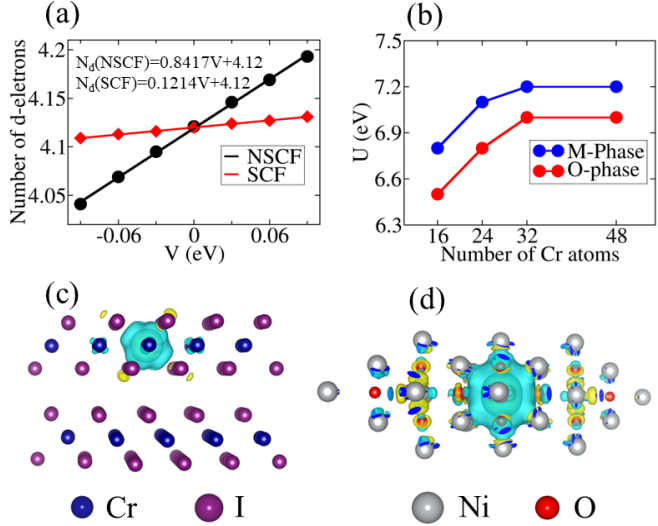

Figure 2: (a) Variation of d electrons number with the small additional potential in Cr-3d orbitals in O phase CrI$_2$, (b) Convergence of $U$ with supercell size. (c, d) Difference in response charge distribution in CrI$_2$ (c) compared to that in NiO (d).

### 3.3. Thermodynamic Stability

The evaluated formation energy of $CrI_2$, -3.67 eV/f.u. (f.u. stands for formula unit), is much larger than the calculated formation energy of $CrI_3$, -5.38 eV/f.u., indicating the low stability of the compound. Accordingly, the chemical potential diagram (Fig. 3) shows stringent requirements for the formation. The formation of $CrI_2$ requires $\mu_{Cr} + 2\mu_I \geq \Delta H_f(CrI_2)$, $\mu_{Cr} + 3\mu_I \leq \Delta H_f(CrI_3)$, $\mu_{Cr} \leq 0$ and $\mu_I \leq 0$ where $\mu$ represents chemical potentials and $\Delta H_f$ represents the heat of formations. As a result, $CrI_2$ forms only in a very narrow range of chemical potentials (0 eV $< \mu_{Cr} < -0.25$ eV; and $-1.78$ eV $< \mu_I < -1.84$ eV) at the bottom right corner of Fig. 3 (enlarged in the inset). This finding is consistent with the preparation reaction, $CrI_3(s) \rightarrow CrI_2(s) + \frac{1}{2}I_2(g)$ [9], and high temperature powder fusion[11] method used to grow $CrI_2$.

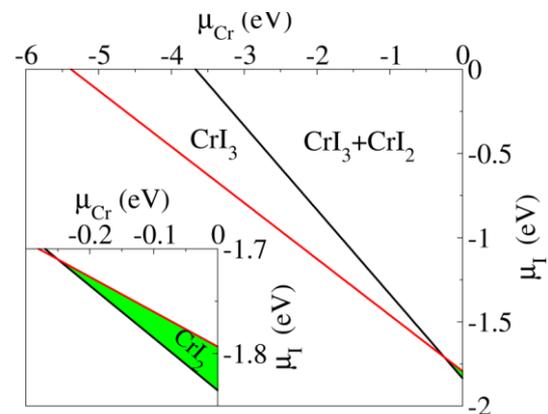

Figure 3: Cr-I phase diagram. $CrI_2$ is stable only in a small region at the bottom right corner; the region is enlarged in the inset.

In comparison, the range for $CrI_3$ is much wider 0 eV $< \mu_{Cr} < -5.38$ eV and 0 eV $< \mu_I < -1.78$ eV. As a result, $CrI_3$ is much more prevalent than $CrI_2$.

The $CrI_2$ monolayer can be easily cleaved from the bulk and shows no signs of instability. The cleavage is measured by the energy difference between the fractured structure and equilibrium ground state as a function of inter-layer distance, $d$. The cleavage converges to 0.25 J/m² (Fig. S2 (b)) similar to that for graphite ( experimental[34]: 0.36 J/m², DFT[35]: 0.43 J/m² ), $CrI_3$ DFT[36]: 0.31 J/m²) and other 2D materials[37]. Imaginary phonon modes associated with the lattice instability are absent in the phonon band structure in the entire Brillouin zone (Fig. S2(c)). Room temperature molecular dynamics simulation running over 10 ps does not show a sign of the phase change – evaluated total energy of the system remains free of discontinuity (Fig. S2(d)).

### 3.4. Electric Polarization

The ground state *O* phase $CrI_2$ is non-centrosymmetric and lacks mirror symmetry along the *z*-axis. Such a structure allows for the presence of polarization. Both bulk and two layers of O-phase with vacuum in the c-axis (i.e. bilayer) can have finite polarization. Berry phase calculation, based on the DFT results in a small polarization along the *z*-axis of 0.157 $\mu C/cm^2$ for the bulk crystal and 0.034 $\mu C/cm^2$ for bilayer $CrI_2$, which are comparable to the 0.70 $\mu C/cm^2$ of bulk $In_2Se_3$[38] and 0.051 $\mu C/cm^2$ of bilayer $WTe_2$[39]. However, these values are smaller compared to perovskite-oxide ferroelectric materials like $BaTiO_3$, where our calculated polarization of 41.3 $\mu C/cm^2$ aligns closely with the literature-reported value [40] of 47 $\mu C/cm^2$ and 7.6 $\mu C/cm^2$ of 2D antiferroelectric hybrid perovskite.[41]

Fig. 4 (a) shows the structure of polarization up ($P \uparrow$) and polarization down ($P \downarrow$) states of O phase CrI$_2$ along with the intermediate state. The transition from $P \uparrow$ to $P \downarrow$ can be realized by sliding the top layer CrI$_2$ by 0.308b along the $-\hat{b}$ direction. This is equivalent to changing the stacking from Cr-I to I-Cr which can be obtained by the mirror operation, $M_z$, similar to a process of changing M-X to X-M stacking in bilayer transition metal dichalcogenides MX$_2$ [42], and other 2D materials [38,39,42–46]. A relatively small energy barrier, of 0.104 eV/f.u (Fig. 4(b)), separates the two polarization states, a value reasonable for dynamic switching. For comparison, an energy barrier of 0.066 eV/f.u. is reported for In$_2$Se$_3$.[38] The ferroelectric transition temperature based on the mean-field approximation, evaluates to be $T_c \sim \frac{2\Delta}{3k_B} = 444$ K, where $k_B$ is the Boltzmann constant, and $\Delta$ is an energy barrier, thus showing stable polar ordering at the room temperature. These transition can happen with external perturbations including pressure and infra-red optical pulses as observed in MoS$_2$.[47]

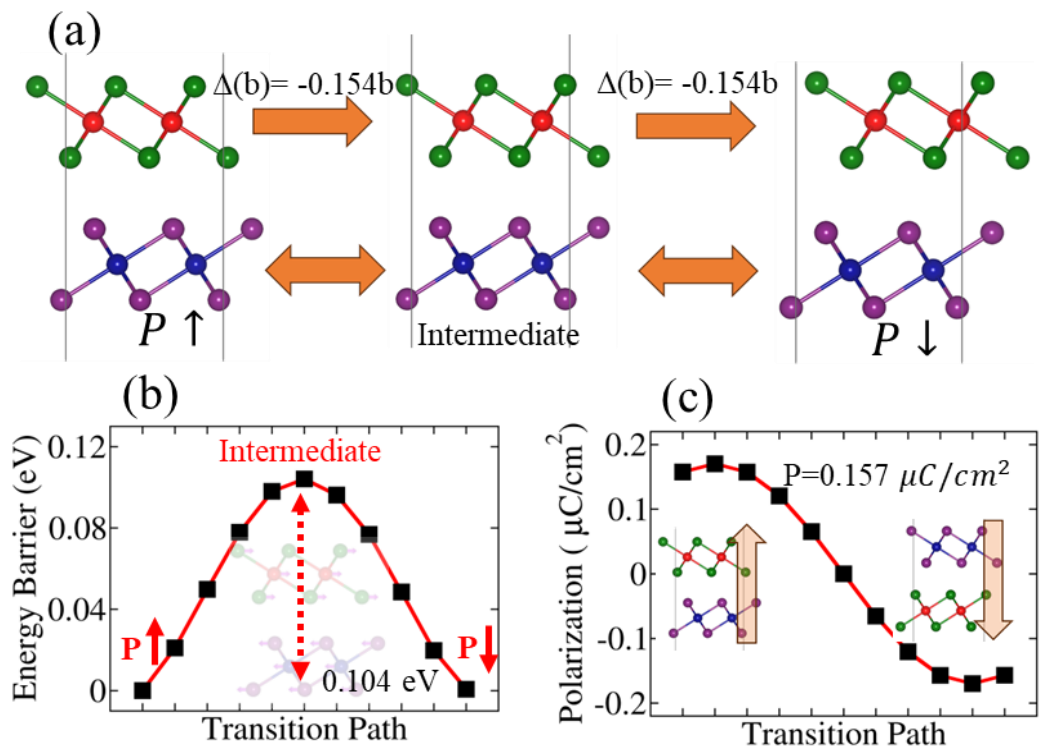

Figure 4: (a) Crystal structure viewed from the a-axis showing polarization switching mechanism with the interlayer sliding. (b) Energy for the structures along the transition path. (c) Variation of polarization along the transition path.

## 3.5. Collinear Magnetic Phase

By comparing the energy of the *M* and *O* structural phases and all possible magnetic phases (Fig. S1) of CrI$_2$ using the $U$ determined above (Table S3), we find that FM *O* phase CrI$_2$ to be the ground state which is in agreement with the previous report[7]. The energy difference between magnetic phases: FM, and different AFM phases, G-type (AFM-III), A-type (AFM-I), and C-type (AFM-II) (Fig. S1), however, is small (< 25 meV) suggesting the possibility of mixed magnetic phases at ambient conditions. The ground state phase as reported earlier depends on the value of $U$. While the orthorhombic *O* phase stays ground state for all values of $U$ tested (0 − 10 eV), the magnetic phase changes depending on the value of $U$. The AFM phase gains energy upon lowering $U$ values and when $U$ < 6.0 eV, the AFM becomes ground states in both *M* and *O* phase (Table S3).

Magnetic interactions responsible for stabilizing the ground state are estimated by mapping DFT energy to the Heisenberg Hamiltonian, $H = -\sum_{\langle ij \rangle} J_1 \hat{s}_i \cdot \hat{s}_j |s|^2 - \sum_{\langle ij \rangle} J_2 \hat{s}_i \cdot \hat{s}_j |s|^2 + E_{\text{others}}$. Here, $s_i$ is the direction of spin of atom $i$, $|s|$ is the value of spin, $J_1$ and $J_2$ represent the first and second nearest neighbor interactions, and $E_{\text{others}}$ are other energy contributions that are expected to remain the same in different magnetic phases. To calculate $J_1$ and $J_2$, we evaluate the energy for different magnetic configurations of CrI$_2$: FM, Néel type AFM (AFM-Neel), and zigzag type AFM (AFM-zigzag) (Fig. 5(a)). Their energies can be expressed as: $E_{FM} = E_{others} - (2J_1 + 4J_2)$, $E_{AF\text{-}Néel} = E_{others} - (2J_1 - 4J_2)$, and $E_{AF\text{-}Zigzag} = E_{others} + (2J_1)$. Solving these equations simultaneously, we find $J_1$ and $J_2$ are both FM with values of 4.35 meV and 0.94 meV, which are slightly smaller than $J_1 = 6.44$ meV and $J_2 = 1.44$ meV in CrI$_3$[36]. Fig. 5(b) shows the magnetic phase diagram obtained by varying $J_1$ and $J_2$, possibly with the external perturbation including bi-axial strain. The FM nature of $J_1$ and $J_2$ originates from strong 90° super exchange interaction between Cr $d$ −orbitals through I−$p$ orbitals as dictated by the Goodenough-Kanamori-Anderson rule[48]. The direct exchange between Cr-Ions, which is typically weakly AFM, is rather weak in this case, because of the large separation as shown in Fig. 5(c).

Magnetic anisotropy of CrI$_2$ is calculated. Fig. 5(d) shows a typical cosine energy curve of uniaxial anisotropy with an easy axis along $c$ axis. The magnetic anisotropy energy (MAE) is 0.58 meV, which is comparable to the MAE of 0.65 meV of CrI$_3$ [49]. Large anisotropy ensures the existence of magnetism in the 2D limit despite large thermal fluctuations.[50] The ferromagnetic transition temperature is estimated to be 42 K for bulk and 32 K in monolayers, based on the Monte-Carlo simulation[51] using Heisenberg Hamiltonian with $J_1$ and $J_2$ as well as the magnetization dynamics calculation based on solution of Landau-Lifshitz-Gilbert (LLG) equations (Fig. S3).

The FM phase is robust with respect to biaxial strain. Fig. 5(e) and Fig. 5(f) shows evolution of exchange parameters $J_1$ and $J_2$ as a function of biaxial strain. The $J_1$ and $J_2$ are calculated similarly to that of the unstrained condition by mapping DFT energy to the Heisenberg Hamiltonian. Fig. 5(g) shows the strain-induced magnetic phase diagram. As can be seen from the figure, the dominant magnetic interaction $J_1$ remains positive until a strain of −6%, $J_2$ is a bit more sensitive, but its size is about 4 times smaller than $J_1$. The observation is consistent with the fact that 90° AFM direct exchange dominates FM super exchange when inter-ion distance decreases. The exchange interaction can also be modulated by electrostatic doping[36].

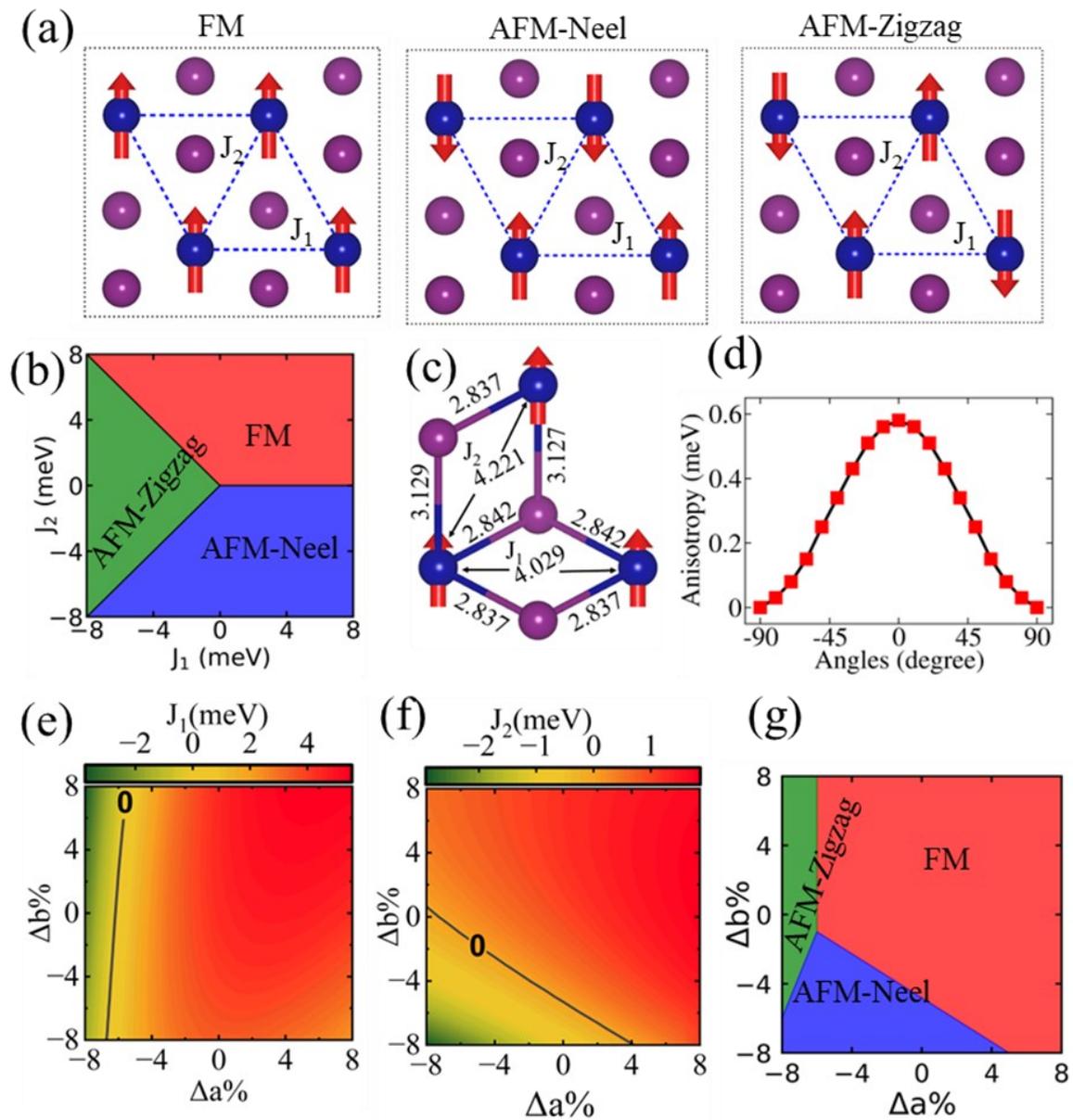

Figure 5: (a) Different co-linear magnetic configurations: FM, Néel-type AFM (AFM-Néel), and Zigzag-type AFM (Zigzag-AFM) considered for calculating the exchange parameters. (b) Magnetic phase diagram deduced based on exchange coefficient $J_1$ and $J_2$. (c) Cr-Cr distances involved in direct and super exchange interaction. (d) Variation of magnetic anisotropy with the angles, 0°, and 90° corresponds to the easy axis and hard axis in the crystal. (e,f) Dependence of (e) $J_1$ and (f) $J_2$ with the in-plane lattice strain in monolayer $CrI_2$. The black line indicates the boundaries of positive and negative values of $J_1$ and $J_2$. (g) Magnetic phase diagram under the strain along $a$ and $b$ direction. The different color indicates the different magnetic orderings.

## 3.6. Magnetic Skyrmions

Non-centrosymmetric structure enables nonzero DMI. The bilayer/bulk O phase CrI$_2$ has switchable polarization with inversion symmetry breaking. According to Moriya's symmetry rules,[16] in the bilayer CrI$_2$, the DMI vector $d_{ij}$ between pair of Cr atoms $i$ and $j$, with localized moment can be written as, $d_{ij} = d^N(\mathbf{r}_{ij} \times \hat{\mathbf{n}})$, where $d^N$ represents the strength of Néel type DMI. The $\mathbf{r}_{ij}$ is the distance vector between atoms $i$ and $j$ and direction $\hat{\mathbf{n}}$ is perpendicular to $\mathbf{r}_{ij}$. In the bilayer CrI$_2$, the Néel type DMI is contributed by the super exchange interaction between Cr atoms through intermediate I atom.[52] In this mechanism, $\hat{\mathbf{n}} \propto \mathbf{r}_{ik} - \mathbf{r}_{jk}$ where $\mathbf{r}_{ik}$ is the displacement between Cr atoms $i$ and $j$ and intermediate I atom $k$. The DMI vector is perpendicular to the plane formed by Cr-I-Cr superexchange path. The *M* phase CrI$_2$ has no net DMI due to the inversion symmetry. In the *O* phase CrI$_2$, the broken inversion symmetry causes the DMI vector mediated by the upper I atom to no longer cancel out the one mediated by the lower I atom. This is due to the different environment between upper and lower interface and the charge redistribution from the electric polarization. Consequently, non-zero DMI vectors could emerge.

The DMI is evaluated by rotating localized spin at Cr ions along the *xz* plane in CW and ACW directions as shown in Fig. S4 with detail derivation in S6, which is successfully applied to metallic interface[53] as well as 2D magnetic system[54]. The constrained noncolinear DFT method incorporates a penalty term in the total energy to align the local magnetic moment in a specified direction. The energy difference between these two configurations, known as DMI energy, is -0.022 meV/f.u. and 0.031 meV/f.u for $P \uparrow$ and $P \downarrow$ configurations as shown in Fig. 6(a). The sign change of DMI with the polarization reversal is determined by the symmetry operation $M_z$ between $P \uparrow$ and $P \downarrow$ states.

The DMI in the material, along with Heisenberg interaction and anisotropy, can induce the formation of skyrmions. We used micromagnetic simulations for the analysis of skyrmions. In the micromagnetic simulation, the dynamics of magnetic moment, $m$ defined as $M/M_s$, under external magnetic field, $H$, is evaluated using Landau-Lifshitz Gilbert (LLG) equations, $\frac{dm}{dt} = \frac{\gamma_0}{1+\alpha^2} m \times H_{eff} - \frac{\gamma_0 \alpha^2}{1+\alpha^2} m \times (m \times H_{eff})$, where $H_{eff} = -\frac{1}{\mu_0 M_s}\frac{\delta w(m)}{\delta m}$. The energy density defined as $w(m) = 1/V \int_V [-\mu_0 m \cdot H - Am \cdot \nabla^2 m - K(m \cdot u)^2 + Dm \cdot (\nabla \times m)]dV$ contains the exchange stiffness, $A$ the anisotropy constant, $K$ and the DMI constants, $D$ that relates to DFT evaluated exchange constant, $J$, the anisotropy constant, $k$ and the DMI energy, $d$. Their estimated values are: A = 0.85 pJ/m, K = 0.91 MJ/m³, D (P↑) = -0.0045 mJ/m², D (P↓) = 0.0036 mJ/m³ and m = 3.6×10⁵ A/m. The demagnetization effect is included in anisotropy energy using effective anisotropy[55] $K_e = K - \frac{\mu_0 M_s^2}{2}$ in micromagnetic simulation. (Details in supplementary materials S7) The simulation however did not show existence of stable skyrmions, which is consistent with the fact that estimated skyrmions radius[55], $R = \pi D \sqrt{\frac{A}{16AK_e^2 - \pi^2 K_e D^2}}$ =0.4

Å is too small required for spin rotations. Radius of at least few-atomic distance is a necessity for spin rotations.

The larger DMI and smaller magnetic anisotropy are needed to form stable skyrmions. Earlier, different strategies including doping[56], applying an electric field [57–59], or intercalating the atoms between the layers[60,61] are suggested to enhance the DMI. Here, we used intercalation of 5d transition metal Pt in the center between two layers of $CrI_2$ to enhance DMI. With Pt intercalation, as shown in Fig. S7, in Pt-4($CrI_2$), there is a significant reduction in out-of-plane anisotropy to 0.14 meV from 0.58 meV and enhancement of DMI energy to 0.28 meV/f.u. form 0.022 meV/f.u because of the high SOC of Pt. The absence of imaginary phonon modes in the phonon band structure, as shown in Fig S7(c, d), confirms the stability of the intercalated structure. The intercalation enhances ferroelectricity of 0.63 $\mu C/cm^2$ and the DMI is switchable according to the ferroelectric polarization. The obtained values of DMI and anisotropy are comparable to 0.44 meV and 0.104 meV observed for Co-intercalated $MoSe_2$.[60] All atomistic and micromagnetic values are listed in Tables S4 (supplemental Materials).

Fig. 6(b, c) shows skyrmions evolved from randomly distributed spins in a 2D mesh during micromagnetic simulations under the application of a small external field $H = 0.05T$ in two different polarization states. The topological charge, $Q = \frac{1}{4\pi} \int m \cdot \left(\frac{\partial m}{\partial x} \times \frac{\partial m}{\partial y}\right) \partial x \, \partial y$ evaluates to be -1 for both cases confirming the emergence of skyrmions. Owing to the opposite sign of DMI in different polarization states, handedness of emergent Néel type skyrmions are opposite as shown in Fig. 6.

With the new values of the magnetic parameter for Pt-4($CrI_2$), the estimated radius is 9.4 nm, closely matching the simulated radius of ~10 nm in the absence of a magnetic field. Fig. S8 shows the need for small magnetic fields to resolve distinct skyrmions. When increasing magnetic field strength from 0.05T to 0.1T, skyrmion size decreases but their number increases, as reported earlier in $CrI_3$[58] and prototypical metallic systems Ir (111) and FePd [62]. As the magnetic field increases to a higher value (0.5T), all the skyrmions vanish as expected since DMI would not be able to preserve the magnetic texture anymore.

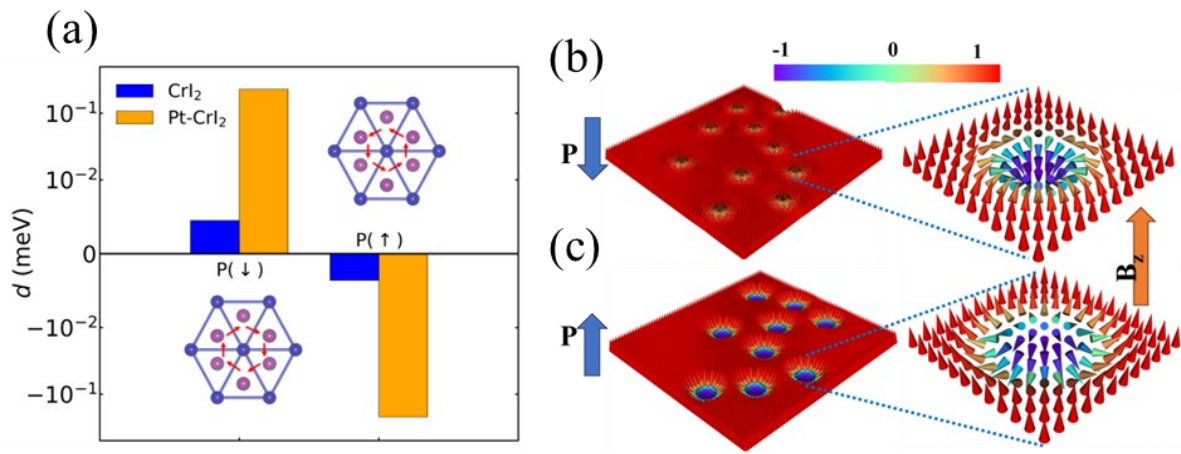

Figure 6: DMI energy of CrI$_2$ bilayer and Pt-intercalated CrI$_2$ bilayer for $P \downarrow$ and $P \uparrow$ (a). Magnetic skyrmions under a magnetic field of 0.05 T for $P \downarrow$(b) and $P \uparrow$(c). The right panel of (b) and (c) shows individual skyrmions. The arrow colors indicate the z-component of magnetization, $m_z$, as shown in the color bar.

## 4. Summary

Using First-principles DFT calculations we show that CrI$_2$ is a multiferroic van der Waals crystal in its pristine form that can be easily exfoliated. The exfoliated layers are dynamically and thermally stable even in their single-layer limit. We show that shifting the van der Waals layer along *b* axis of the ground state orthorhombic phase changes out-of-plane polarization. The magnetic exchange interactions with the explicitly calculated value of Hubbard U predict ferromagnetism, with a large magnetic moment of 3.9 $\mu_B$/Cr, large magnetic anisotropy, 0.58 meV/Cr. However, the ferromagnetic competes closely with other anti-ferromagnetic phases, and by altering exchange interaction by external means, e.g., strain other magnetic phases may arise.

The absence of inversion symmetry in the O-phase, bulk, or bilayer structure of the crystal gives rise to a non-zero DMI, enabling the formation of skyrmions whose chirality can be controlled by ferroelectric polarization in CrI$_2$. Our work thus brings the possible usage of CrI$_2$ in the creation and manipulation of skyrmions that may be useful in spintronics memory devices, including racetrack memories[63].

**Acknowledgment**

The work at South Dakota Mines and Technology is supported by the Emergent Quantum Materials and Technologies (EQUATE) Award OIA-2329159. The research work at the University of Nebraska-Lincoln (theoretical analysis of DMI) is supported by grant number DE-SC0023140, funded by the U.S. Department of Energy, Office of Science, Basic Energy Sciences (K.H. and E. Y. T.). Computations were performed using the Holland Computing Center at the University of Nebraska–Lincoln, the High-Performance Computing Center at South Dakota State University, and the Lawrence supercomputer at the University of South Dakota.

**References**

1. Huang, B. *et al*. Layer-dependent ferromagnetism in a van der Waals crystal down to the monolayer limit. *Nature* **546**, 270–273 (2017).
2. Sivadas, N., Okamoto, S., Xu, X., Fennie, C. J. & Xiao, D. Stacking-Dependent Magnetism in Bilayer CrI 3. *Nano Lett.* **18**, 7658–7664 (2018).
3. Jiang, P. *et al*. Stacking tunable interlayer magnetism in bilayer CrI3. *Phys. Rev. B* **99**, 1–9 (2018).
4. Paudel, T. R. & Tsymbal, E. Y. Spin Filtering in CrI 3 Tunnel Junctions. *ACS Appl. Mater. Interfaces* **11**, 15781–15787 (2019).
5. Marian, D., Soriano, D., Cannavó, E., Marin, E. G. & Fiori, G. Electrically tunable lateral spin-valve transistor based on bilayer CrI3. *npj 2D Mater. Appl.* **7**, 1–7 (2023).

6. Besrest, F. & Jaulmes, S. Structure cristalline de l'iodure de chrome, CrI2. *Acta Crystallogr. Sect. B Struct. Crystallogr. Cryst. Chem.* **29,** 1560–1563 (1973).
7. Peng, L. et al. Mott phase in a van der Waals transition-metal halide at single-layer limit. *Phys. Rev. Res.* **2,** 023264 (2020).
8. Cai, X., Xu, Z., Ji, S. H., Li, N. & Chen, X. Molecular beam epitaxy growth of iodide thin films. *Chinese Phys. B* **30,** 028102 (2021).
9. Handy, L. L. & Gregory, N. W. A Study of the Chromous—Chromic Iodide Equilibrium. *J. Am. Chem. Soc.* **72,** 5049–5051 (1950).
10. Zhang, S., Tang, F., Song, X. & Zhang, X. Structural phase transitions and Raman identifications of the layered van der Waals magnet CrI2. *Phys. Rev. B* **105,** 104105 (2022).
11. Schneeloch, J. A., Liu, S., Balachandran, P. V., Zhang, Q. & Louca, D. Helimagnetism in the candidate ferroelectric CrI2. *Phys. Rev. B* **109,** 144403 (2024).
12. Zhao, Y., Liu, H., Gao, J. & Zhao, J. Transition of CrI2 from a two-dimensional network to one-dimensional chain at the monolayer limit. *Phys. Chem. Chem. Phys.* **23,** 25291–25297 (2021).
13. Kulish, V. V. & Huang, W. Single-layer metal halides MX2 (X = Cl, Br, I): Stability and tunable magnetism from first principles and Monte Carlo simulations. *J. Mater. Chem. C* **5,** 8734–8741 (2017).
14. Yang, L., Gao, Y., Wu, M. & Jena, P. Interfacial triferroicity in monolayer chromium dihalide. *Phys. Rev. B* **105,** 094101 (2022).
15. Bo, X., Fu, L., Wan, X., Li, S. & Pu, Y. Magnetic structure and exchange interactions of transition metal dihalide monolayers: First-principles studies. *Phys. Rev. B* **109,** 014405 (2024).
16. Moriya, T. Anisotropic Superexchange Interaction and Weak Ferromagnetism. *Phys. Rev.* **120,** 91–98 (1960).
17. Mühlbauer, S. et al. Skyrmion Lattice in a Chiral Magnet. *Science.* **323.**5916, 915–919 (2009).
18. Bogdanov, A. N. & Yablonskii, D. A. Thermodynamically stable 'vortices' in magnetically ordered crystals. The mixed state of magnets. *Zh. Eksp. Teor. Fiz* **95,** 178 (1989).
19. Kresse, G. & Furthmüller, J. Efficient iterative schemes for ab initio total-energy calculations using a plane-wave basis set. *Phys. Rev. B* **54,** 11169–11186 (1996).
20. Kresse, G., Joubert, D., Kresse, G. & Joubert, D. From ultrasoft pseudopotentials to the projector augmented-wave method. *Phys. Rev. B* **59,** 1758–1775 (1999).
21. Blöchl, P. E. Projector augmented-wave method. *Phys. Rev. B* **50,** 17953–17979 (1994).
22. Hendrik J. Monkhorst. Special points fro Brillouin-zone integretions. *Phys. Rev. B* **13,** 5188–5192 (1976).
23. Dudarev, S. L., Botton, G. A., Savrasov, S. Y., Humphreys, C. J. & Sutton, A. P. Electron-energy-loss spectra and the structural stability of nickel oxide: An LSDA+U study. *Phys. Rev. B* **57,** 1505–1509 (1998).
24. Cococcioni, M. & de Gironcoli, S. Linear response approach to the calculation of the effective interaction parameters in the LDA+U method. *Phys. Rev. B* **71,** 035105 (2005).
25. Grimme, S., Antony, J., Ehrlich, S. & Krieg, H. A consistent and accurate ab initio parametrization of density functional dispersion correction (DFT-D) for the 94 elements H-Pu. *J. Chem. Phys.* **132**, 154104 (2010).
26. Togo, A. & Tanaka, I. First principles phonon calculations in materials science. *Scr. Mater.* **108**, 1–5 (2015).
27. Beg, M., Pepper, R. A. & Fangohr, H. User interfaces for computational science: A domain specific language for OOMMF embedded in Python. *AIP Adv.* **7**, (2017).
28. Donahue, M. J. & Porter, D. G. *OOMMF User's Guide, Version 1.0*. (US Department of Commerce, National Institute of Standards and Technology, 1999).
29. Tracy, J. W., Gregory, N. W., Stewart, J. M. & Lingafelter, E. C. The crystal structure of chromium (II) iodide. *Acta Crystallogr.* **15**, 460–463 (1962).
30. Function, C., Ridge, O., Ridge, O. & Jr, R. F. Structure Cristalline de l'Iodure de Chrome: CrI2. 1560–1563 (1972).
31. Chen, S. *et al.* Boosting the Curie Temperature of Two-Dimensional Semiconducting CrI3 Monolayer through van der Waals Heterostructures. *J. Phys. Chem. C* **123**, 17987–17993 (2019).
32. Padmanabhan, P. *et al.* Coherent helicity-dependent spin-phonon oscillations in the ferromagnetic van der Waals crystal CrI3. *Nat. Commun.* **13**, 1–8 (2022).
33. Soriano, D., Rudenko, A. N., Katsnelson, M. I. & Rösner, M. Environmental screening and ligand-field effects to magnetism in CrI3 monolayer. *npj Comput. Mater.* **7**, 1–10 (2021).
34. Wang, W. *et al.* Measurement of the cleavage energy of graphite. *Nat. Commun.* **6**, 1–7 (2015).
35. Björkman, T., Gulans, A., Krasheninnikov, A. V. & Nieminen, R. M. van der Waals Bonding in Layered Compounds from Advanced Density-Functional First-Principles Calculations. *Phys. Rev. Lett.* **108**, 235502 (2012).
36. Zhang, W. B., Qu, Q., Zhu, P. & Lam, C. H. Robust intrinsic ferromagnetism and half semiconductivity in stable two-dimensional single-layer chromium trihalides. *J. Mater. Chem. C* **3**, 12457–12468 (2015).
37. Lipatov, A. *et al.* Electronic transport and polarization-dependent photoresponse in few-layered hafnium trisulfide (HfS3) nanoribbons. *J. Mater. Chem. C* **11**, 9425–9437 (2023).
38. Ding, W. *et al.* Prediction of intrinsic two-dimensional ferroelectrics in In 2 Se 3 and other III 2 -VI 3 van der Waals materials. *Nat. Commun.* **8**, 1–8 (2017).
39. Yang, Q., Wu, M. & Li, J. Origin of Two-Dimensional Vertical Ferroelectricity in WTe 2 Bilayer and Multilayer. *J. Phys. Chem. Lett.* **9**, 7160–7164 (2018).
40. Zhang, Y., Sun, J., Perdew, J. P. & Wu, X. Comparative first-principles studies of prototypical ferroelectric materials by LDA, GGA, and SCAN meta-GGA. *Phys. Rev. B* **96**, 035143 (2017).
41. Li, L. *et al.* 2D Antiferroelectric Hybrid Perovskite with a Large Breakdown Electric Field And Energy Storage Density. *Adv. Funct. Mater.* **33**, 2305524 (2023).
42. Vizner Stern, M. *et al.* Interfacial ferroelectricity by van der Waals sliding. *Science* **372**.6549 1462–1466 (2021).
43. Guan, Z. *et al.* Recent Progress in Two-Dimensional Ferroelectric Materials. *Adv. Electron. Mater.* **6**, 1–30 (2020).
44. Wang, X. *et al.* Interfacial ferroelectricity in rhombohedral-stacked bilayer transition metal dichalcogenides. *Nat. Nanotechnol.* **17**, 367–371 (2022).
45. Li, L. & Wu, M. Binary Compound Bilayer and Multilayer with Vertical Polarizations: Two-Dimensional Ferroelectrics, Multiferroics, and Nanogenerators. *ACS Nano* **11**,

6382–6388 (2017).
46. Lu, H. *et al.* 3D Domain Arrangement in van der Waals Ferroelectric α-In2Se3. *Adv. Funct. Mater.* **2403537**, 1–8 (2024).
47. Park, J., Yeu, I. W., Han, G., Hwang, C. S. & Choi, J. H. Ferroelectric switching in bilayer 3R MoS2 via interlayer shear mode driven by nonlinear phononics. *Sci. Rep.* **9**, 1–9 (2019).
48. Goodenough, J. B. *Magnetism and the Chemical Bond*. (Hassell Street Press, 1963).
49. Lado, J. L. & Fernández-Rossier, J. On the origin of magnetic anisotropy in two dimensional CrI 3. *2D Mater.* **4**, 035002 (2017).
50. Mermin, N. D. & Wagner, H. Absence of ferromagnetism or antiferromagnetism in one-or two-dimensional isotropic Heisenberg models. *Phys. Rev. Lett.* **17**, 1133 (1966).
51. Evans, R. F. L. *et al.* Atomistic spin model simulations of magnetic nanomaterials. *J. Phys. Condens. Matter* **26**, (2014).
52. Bogdanov, A. & Hubert, A. Thermodynamically stable magnetic vortex states in magnetic crystals. *J. Magn. Magn. Mater.* **138**, 255–269 (1994).
53. Yang, H., Thiaville, A., Rohart, S., Fert, A. & Chshiev, M. Anatomy of Dzyaloshinskii-Moriya Interaction at Co/Pt Interfaces. *Phys. Rev. Lett.* **115**, 267210 (2015).
54. Yang, H., Liang, J. & Cui, Q. First-principles calculations for Dzyaloshinskii–Moriya interaction. *Nat. Rev. Phys.* **5**, 43–61 (2022).
55. Wang, X. S., Yuan, H. Y. & Wang, X. R. A theory on skyrmion size. *Commun. Phys.* **1**, 1–7 (2018).
56. Huang, K., Schwartz, E., Shao, D.-F., Kovalev, A. A. & Tsymbal, E. Y. Magnetic antiskyrmions in two-dimensional van der Waals magnets engineered by layer stacking. *Phys. Rev. B* **109**, 024426 (2024).
57. Liu, J., Shi, M., Lu, J. & Anantram, M. P. Analysis of electrical-field-dependent Dzyaloshinskii-Moriya interaction and magnetocrystalline anisotropy in a two-dimensional ferromagnetic monolayer. *Phys. Rev. B* **97**, 054416 (2018).
58. Behera, A. K., Chowdhury, S. & Das, S. R. Magnetic skyrmions in atomic thin CrI3 monolayer. *Appl. Phys. Lett.* **114**, (2019).
59. Srivastava, T. *et al.* Large-Voltage Tuning of Dzyaloshinskii-Moriya Interactions: A Route toward Dynamic Control of Skyrmion Chirality. *Nano Lett.* **18**, 4871–4877 (2018).
60. Shao, Z. *et al.* Multiferroic materials based on transition-metal dichalcogenides: Potential platform for reversible control of Dzyaloshinskii-Moriya interaction and skyrmion via electric field. *Phys. Rev. B* **105**, 174404 (2022).
61. Zheng, G. *et al.* Tailoring Dzyaloshinskii–Moriya interaction in a transition metal dichalcogenide by dual-intercalation. *Nat. Commun.* **12**, 1–7 (2021).
62. Romming, N. *et al.* Writing and Deleting Single Magnetic Skyrmions. *Science*. **341**.6146, 636–639 (2013).
63. Fert, A., Cros, V. & Sampaio, J. Skyrmions on the track. *Nat. Nanotechnol.* **8**, 152–156 (2013).